# Ordered arrays of InGaN/GaN dot-in-a-wire nanostructures as single photon emitters

Snežana Lazić*[a], Ekaterina Chernysheva[a], Žarko Gačević[b], Noemi García-Lepetit[a], Herko P. van der Meulen[a], Marcus Müller[c], Frank Bertram[c], Peter Veit[c], Jürgen Christen[c], Almudena Torres-Pardo[d], José M. González Calbet[d], Enrique Calleja[b], José M. Calleja[a]

[a]Departamento de Física de Materiales, Instituto "Nicolás Cabrera" and Instituto de Física de Materia Condensada (IFIMAC), Universidad Autónoma de Madrid, 28049 Madrid, Spain; [b]ISOM-DIE, Universidad Politécnica de Madrid, 28040 Madrid, Spain; [c]Institute of Experimental Physics, Otto-von-Guericke-University Magdeburg, 39106 Magdeburg, Germany; [d]Centro Nacional de Microscopía Electrónica, Universidad Complutense de Madrid, 28040 Madrid, Spain

## ABSTRACT

The realization of reliable single photon emitters operating at high temperature and located at predetermined positions still presents a major challenge for the development of solid-state systems for quantum light applications. We demonstrate single-photon emission from two-dimensional ordered arrays of GaN nanowires containing InGaN nano-disks. The structures were fabricated by molecular beam epitaxy on (0001) GaN-on-sapphire templates patterned with nanohole masks prepared by colloidal lithography. Low-temperature cathodoluminescence measurements reveal the spatial distribution of light emitted from a single nanowire heterostructure. The emission originating from the topmost part of the InGaN regions covers the blue-to-green spectral range and shows intense and narrow quantum dot-like photoluminescence lines. These lines exhibit an average linear polarization ratio of 92%. Photon correlation measurements show photon antibunching with a $g^{(2)}(0)$ values well below the 0.5 threshold for single photon emission. The antibunching rate increases linearly with the optical excitation power, extrapolating to the exciton decay rate of ~1 ns$^{-1}$ at vanishing pump power. This value is comparable with the exciton lifetime measured by time-resolved photoluminescence. Fast and efficient single photon emitters with controlled spatial position and strong linear polarization are an important step towards high-speed on-chip quantum information management.

**Keywords:** GaN/InGaN nanowire-quantum dots, single photon source, polarized emission

## 1. INTRODUCTION

The quest for fast, efficient and fully polarized single photon emitters (SPEs) operating at high temperatures and located at precise positions is essential for the development of solid state systems for quantum light applications. Different quantum systems have been actively investigated over the last few decades for the realization of such systems in the solid-state. Significant progress toward highly-efficient SPEs has been demonstrated by quantum dots (QDs) in both narrow- and wide-bandgap material systems. Unlike narrow bandgap semiconductors (such as III-As and III-P) which still face several limitations in operating temperature and scalability, wide-bandgap materials based on group III-nitrides offer an attractive platform for high-temperature operation [1-3] and broad emission range from the deep ultraviolet (6.2 eV for AlN) to telecommunication wavelengths (0.7 eV for InN). In particular, the III-nitride-based quantum emitters operating in the ultraviolet-visible region are well suited for small-scale quantum information circuits and free space communication [4]. Also, the special valence-band structure of III-nitrides favors highly polarized light emission, which is essential for linear optical quantum computing [5] and quantum key distribution [6].

Despite the considerable progress in the field, the precise control of the SPEs location is still a major challenge. Standard techniques to produce self-assembled QDs result in randomly distributed SPE. Another approach, recently reported, is the use of semiconductor micro/nanostructures hosting QDs. Several types of such heterostructures [1,7,8], including ordered arrays of site-controlled GaN- [3] and InGaN-based [9] quantum emitters, have been fabricated and proved to be effective for the spatial control of SPEs. Among those, the inclusion of quantum emitters into III-nitride nanowires offers additional benefits. The large surface-to-volume ratio and small interface between the nanowire and the underlying material permit a stain-free epitaxial growth on largely mismatched substrates [10]. The better strain relief reduces the formation of threading dislocations and other structural defects, resulting in high structural quality and improved optical performance (no below band-edge absorption and emission). Another favorable characteristic of dot-in-a-nanowire heterostructures is the increase in the light extraction efficiency and directionality [11,12], which allows the development of highly efficient nano-sources even at the single photon level [13]. Also, the absence of a two-dimensional wetting layer in nanowire-QD structures prevents the leakage of carriers from the QD at higher temperatures, thus increasing the device efficiency.

Most of the reported InGaN/GaN structures [3,7-9] for single-photon generation have been grown by metal-organic vapor-phase deposition (MOCVD) which, due to higher growth temperature commonly employed in this technique, hinders the incorporation of high In concentration and, thus, limits the ability for spectral tuning. Here, we demonstrate ordered arrays of site-controlled SPEs grown by molecular beam epitaxy (MBE), with tunable emission wavelength across the entire blue-to-green spectral region. Our structures consist of hexagonal arrays of GaN nanowires with pyramidal tops hosting InGaN nano-disks. The In-rich inclusions inside the InGaN regions are responsible for linearly polarized non-classical light emission. The formation of such QD-like recombination centers is proven by the narrow (sub-meV) emission linewidths in micro-photoluminescence (µ-PL) spectra.

## 2. EXPERIMENTAL DETAILS

### 2.1 Fabrication of InGaN/GaN nanowire heterostructures

InGaN nano-disks embedded in ordered GaN nanowire arrays were grown by plasma-assisted molecular beam epitaxy (PA-MBE) on (0001) GaN-on-sapphire templates [14,15] patterned with titanium nanohole mask prepared by colloidal lithography [16]. The hexagonal lattice of holes (with 280 nm pitch) obtained in this way permitted a selective area growth (SAG) of InGaN/GaN nanowire heterostructures with typical height and diameter of about 500 and 200 nm, respectively. A scanning electron microscopy (SEM) image of the as-grown sample in Fig. 1(a) shows the formation of two-dimensional ordered arrays of nanowires with an average density of ~$1.6\times10^9$ cm$^{-2}$.

The nanowires exhibit hexagonal cross section with lateral facets defined by non-polar m-planes. In case of homoepitaxial SAG growth on c-plane GaN, the growth front is generally formed by semi-polar (r-) and polar (c-) planes, leading to a pyramidal nanowire top profile [16]. This profile determines the shape of the InGaN nano-disks embedded inside the nanowire tips. A scanning transmission electron microscopy (STEM) image of an individual nanowire in Fig. 1(b) confirms the presence InGaN nano-disk with "trapezoidal" lateral cross-section on both the polar nanowire top facet (region A, upper inset of Fig 1(b)) and semi-polar side facets (region B). The smaller thickness of the InGaN nano-disk on the side-facets contributes to the carrier confinement in the central top InGaN region. The high-resolution STEM image (cf. lower inset of Fig 1(b)) shows an atomically smooth, defect-free c-plane InGaN/GaN interface.

Samples with varying vertical thickness (10-30 nm) and indium content of the InGaN nano-disks were obtained by changing the growth parameters (i.e. growth time and substrate temperature). An additional set of samples was prepared for a single nanowire spectroscopy. The wires were mechanically removed from the native substrate and transferred onto a silicon wafer covered with a titanium metal grid patterned by electron beam lithography (Fig. 2(c)). The grid containing 3×3 µm square apertures was used to re-access a targeted wire. In this way, a low density of dispersed nanowires is obtained, allowing easy optical access to a single one (inset in Fig. 2(c)).

### 2.2 Characterization techniques

The optical properties of our InGaN dot-in-a-wire structures were investigated by µ-PL spectroscopy and low-temperature transmission electron microscopy (TEM) combined with cathodoluminescence (CL) spatial mapping. The samples were mounted in a variable temperature continuous helium flow cryostat and excited by a continuous-wave (cw) helium-cadmium (He-Cd) laser operating at 325 and 442 nm. The laser was focused on the sample surface by a 100× microscope objective with a numerical aperture NA=0.73, resulting in a spot size of ~1.5 µm. The distance between the dispersed nanowires was considerably larger than the laser spot, allowing to study the emission of individual nano-disks. The emitted light collected by the same objective was dispersed by a single grating monochromator with a spectral resolution of ~350 µeV and sent to a liquid nitrogen-cooled charge coupled device (CCD) camera. The polarization measurements were performed by setting a half-wave plate together with a fixed linear polarizer in front of the monochromator entrance slit. This arrangement eliminates the effects of the polarization anisotropy of the monochromator grating.

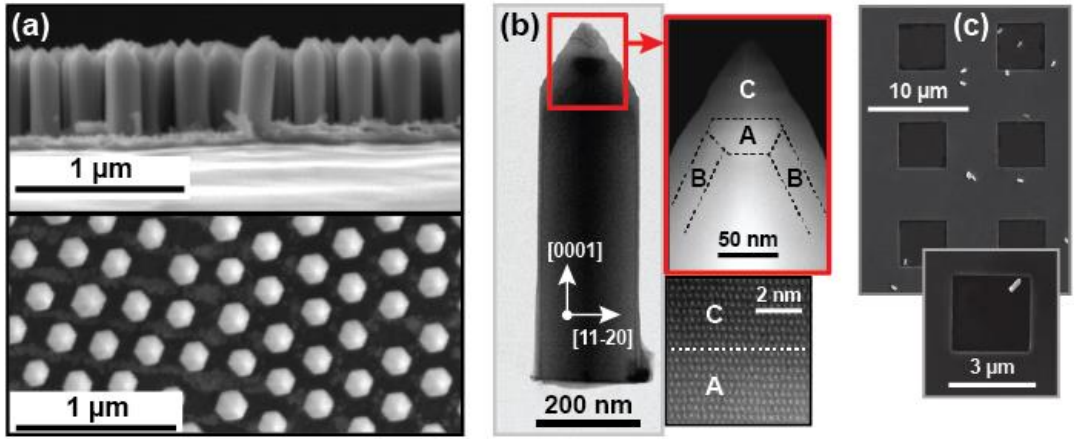

Figure 1. (a) Side and top view SEM image of an as-grown InGaN/GaN nanowire sample. (b) Cross-sectional STEM image of a single nanowire. Upper inset: Magnified view of the InGaN nano-disk embedded in the pyramidal nanowire top. Lower inset: High-resolution STEM image showing atomically smooth InGaN/GaN c-plane interface. (c) SEM image of dispersed nanowires.

The CL detection unit was integrated in a FEI (S) TEM Tecnai F20 equipped with a liquid helium stage (T=10-300 K). The emitted CL is collected by a retractable parabolically-shaped aluminum mirror and directed to the entrance slit of the grating monochromator (MonoCL4, Gatan) equipped with a CCD camera. The material contrast at each point is recorded at the same time as the detection of the CL emission. The acquisition of electrons which are forward-scattered into both high and small solid angles is obtained by a (high-angle) annular dark-field detector from Fischione (model 3000). In order to minimize the sample damage the STEM acceleration voltage was optimized to 80 kV.

A Hanbury-Brown and Twiss (HBT) interferometer [17] placed on the side exit of the µ-PL monochromator was employed for photon correlation measurements. Two single photon counting avalanche photodiodes (APDs) with time resolution $\tau_{IRF}$~350 ps were positioned on the transmission and reflection arms of a 50/50 non-polarizing beam splitter. The detection efficiency of the APDs was ~35% in the measured spectral range. Additionally, a short-pass filter was placed in front of one of the APDs to eliminate optical cross-talk between the two detectors. The signal from the APDs was sent to a time-correlated single photon counting system, which delivers a histogram of the correlation events as a function of the time intervals between photon detection at both detectors. Count rates at the APDs up to $1.5 \cdot 10^5$ photons/s and 32 ps time bin were used to record the correlation histograms.

For time-resolved experiments the same setup with only one APD was used. A pulsed diode laser with 405 nm wavelength, <100 ps pulse width and 40 MHz repetition rate was employed as an excitation source.

## 3. RESULTS AND DISCUSSION

The low-temperature µ-PL measurements on as-grown nanowire samples in Fig. 2(a) reveal three emission bands labeled A, B and C corresponding to two contributions from the InGaN nano-disks and GaN nanowire body, respectively. The lowest energy emission (labeled A) can be tuned across the entire blue-to-green spectral region (~2.85 – 2.35 eV) by varying the growth temperature and the InGaN layer thickness. Moreover, it exhibits sharp and intense PL lines (Fig. 2(b)), which are visible up to 80 K. These narrow peaks are attributed to the presence of QD-like states and, as it will be discussed later, are responsible for the emission of antibunched photons. Their linewidth of 350-650 µeV is comparable to the values reported for high-quality SAG [8] and self-assembled [18,19] InGaN QDs.

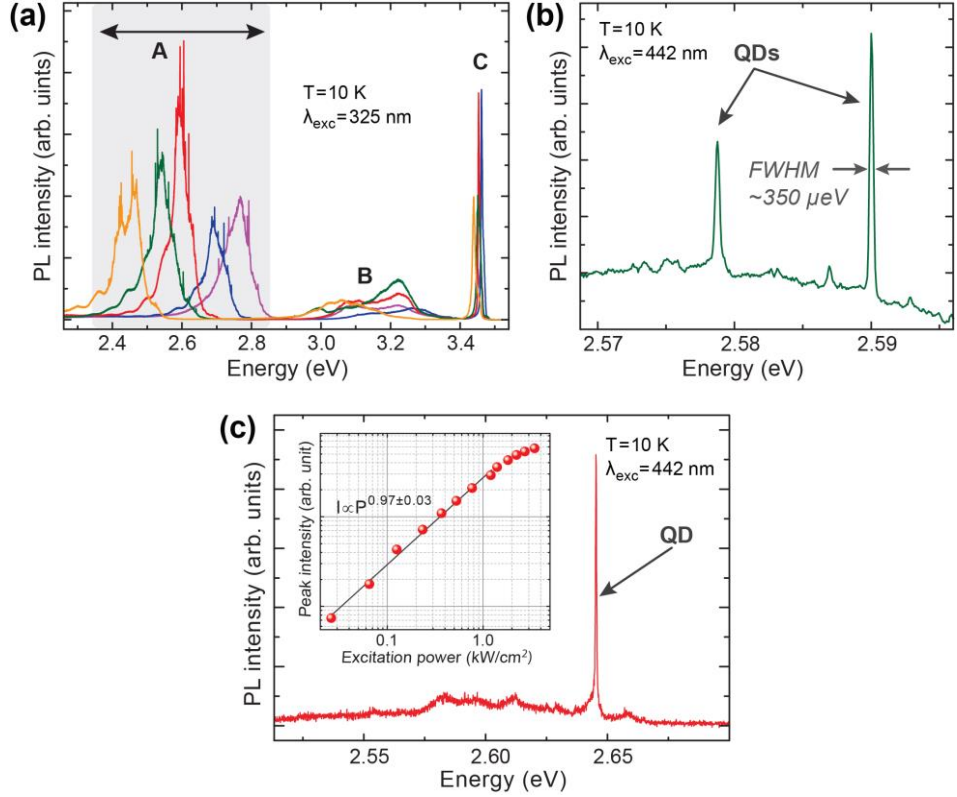

Figure 2. (a) μ-PL spectra measured at 10 K on different as-grown nanowire samples excited at 325 nm. The spectra are normalized to the laser power. (b) PL band labeled A from one of the spectra in (a) centered at 2.54 eV excited with 442 nm laser line. (c) μ-PL spectra recorded at 10 K on a single nanowire from the dispersed batch. Inset: Integrated QD peak intensity as a function of the excitation power density.

Similar spectral distribution is observed in nanowires dispersed on Si wafers. A typical example, shown in Fig. 2(c), presents a dominant PL peak with FWHM of ~500 μeV on top of much weaker background emission. The pump power dependence of the integrated peak intensity (cf. inset in Fig. 2(c)) follows a linear trend, which indicates a ground-state exciton transition. Some of the single nanowires showed more than one well-resolved QD-like peak with linear excitation power dependence. This suggests the presence of more than one emission center in the InGaN region. We expect In-rich clusters formed due to random In fluctuations to be responsible for QD-like confinement potentials. Owing to this fact, the weak background emission in the individual nanowire spectrum centered at ~2.60 eV, is attributed to InGaN region without QD-like confinement.

The polarization-resolved PL spectra were obtained for both as-grown and dispersed nanowires. Figures 3(a) and 3(b) show the dependence of the normalized integrated QD peak intensity $I(\theta)$ on linear polarization angle θ for standing and lying nanowire, respectively. The experimental data were fitted to:

$$I(\theta) = a + b\cos^2(\theta - \theta_0),  \qquad (1)$$

where $\theta_0$ is the linear polarization direction of the QD line. The degree of linear polarization is obtained from the fit using:

$$P = (I_{max} - I_{min})/(I_{max} + I_{min}) = b/(2a+b). \qquad (2)$$

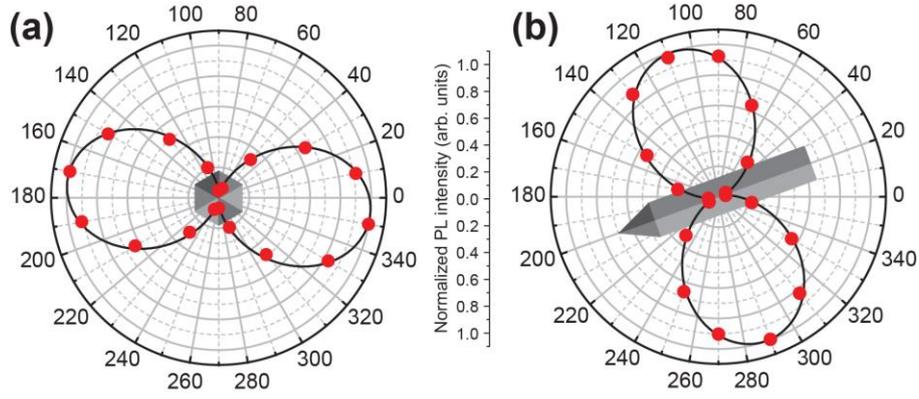

Figure 3. Angular dependence of the integrated QD PL intensity (filled circles) for (a) one of the nanowire QDs from the as-grown sample in Fig. 2(b) and (b) single lying nanowire in Fig. 2. (c). θ=0° is an arbitrary direction. The solid curves are fits to Eq.(1). Insets illustrate the nanowire orientation determined by SEM imaging.

The average polarization degree is found to be ~92%. All emission lines are polarized in the growth plane for all studied nanowires. The observed linear polarization is attributed to the valence band mixing in wurtzite III-nitrides due to the in-plane anisotropy of QD shape as well as internal strain and electric fields [8,18,20,21]. Even a small magnitude of in-plane anisotropy significantly reduces the oscillator strength of one of the cross-polarized transitions, which correspond to degenerate in-plane polarized bright exciton states [20, 21]. This is clearly seen in Fig. 3(a) where the light collected along the nanowire direction (c-axis) has full linear polarization. No indication of preferred crystallographic direction of polarization was found, which suggests a random asymmetry of the QDs.

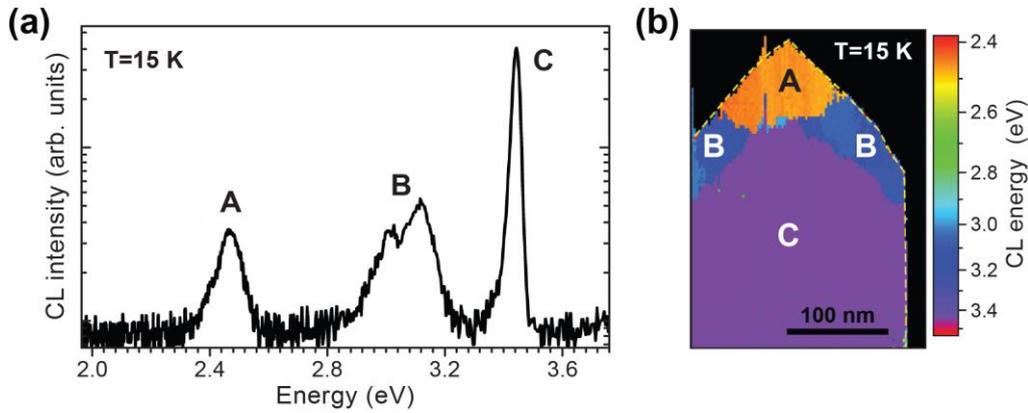

Figure 4. Low temperature CL spectrum of an individual nanowire heterostructure and (b) the corresponding wavelength image. The luminescence bands labeled A, B and C originate from the nanowire tip, side edges of the pyramidal nanowire top and the nanowire body, respectively.

The CL measurements in Fig. 4 reveal the spatial origin of the emission bands observed in the µ-PL spectra. The highest energy luminescence (at ~3.45 eV) originates at the nanowire body and is due to the near-band-edge GaN emission. The one at ~3.06 eV is attributed to the emission from the InGaN sections on semi-polar side-facets (B in Fig. 1(b)), while the one at ~2.45 eV is emitted from the central part of the InGaN nano-disk (A in Fig. 1(b)). The precise origin of the energy separation between the two InGaN contributions is not fully established. It could be explained by different indium concentration, strain and electric field values between the top and side facets of the nanowire apex. Attributing the origin of this energy separation (~0.6 eV) exclusively to In content difference (~20%) [22], we would have In-rich regions located at the topmost part (A) and In-poor regions at the side edges (B) of the InGaN disks. Such compositional variation could be attributed to a higher In incorporation rate on the polar c- compared to the semi-polar r-planes [23]. Moreover, different values of the electric field associated to both spontaneous and piezoelectric polarization at polar and semi-polar planes can also lead to the observed energy shift. Actually, assuming a ~25 nm electron-hole separation, the ~0.6 eV band separation corresponds to a 0.24 MeV/cm electric field difference, which is in good agreement with typical reported values for InGaN/GaN nanostructures [24].

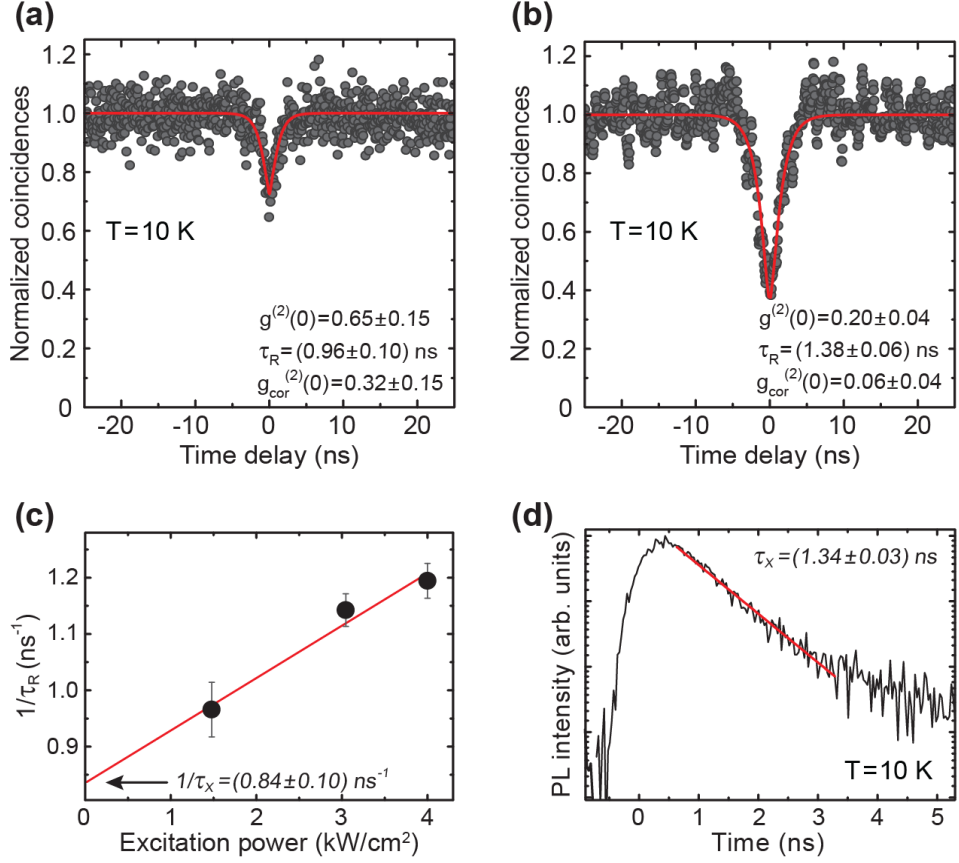

Figure 5. Normalized coincidence counts measured at 10 K from the InGaN apex of the (a) standing and (b) lying nanowires in Fig. 3. The solid lines are fit to the data (circles) by Eq. (3) convoluted Eq. (4), as explained in the text. (c) Antibunching rate (dots) and corresponding linear fit to Eq. (5) (solid line) as a function of the excitation power. (d) Temporal decay of the PL intensity for an individual lying nanowire QD. A mono-exponential fit (solid line) gives a 1.34 ns decay time.

The non-classical character of the QD-like emission from the topmost part of InGaN nano-disks was investigated by HBT interferometry. The normalized coincidence counts histograms for both standing and lying nanowires are depicted in Figs. 5(a) and 5(b). The autocorrelation histograms in Fig. 5 display a pronounced dip at zero time delay ($\tau = 0$). However, the suppression of coincidence counts at $\tau = 0$ is not complete, as expected for an ideal single quantum emitter [25]. The residual counts are due to limited temporal resolution of our HBT setup, as well as background emission originating from neighboring emission centers and dark counts of the HBT detectors. The reduction of coincidence counts at $\tau = 0$ reflects the reduced probability of simultaneous multiple photon emission and, thus, proves the non-classical character of the emitted light. To obtain the actual degree of antibunching, the experimental data are fitted with the standard second-order-auto-correlation-function for continuous-wave excitation:

$$g^{(2)}(\tau) = 1 - \beta \cdot e^{-|\tau/\tau_R|} \quad (3)$$

convoluted with the instrument time response of the HBT detectors [26]:

$$C \cdot e^{-|\tau/\tau_{IRF}|}, \quad (4)$$

(red solid lines in Fig.5 (a) and (b)) where $\beta$ is the antibunching dip value, $\tau$ is the time delay between photons detected at both detectors, $\tau_R$ is the antibunching time constant, $C$ is a normalization factor and $\tau_{IRF}$ is the detectors' temporal resolution. The typical antibunching values derived from these fits ($g^{(2)}(0) = 1 - \beta$) are $0.20 \pm 0.04$ for dispersed individual nanowires (lying) and $0.65 \pm 0.15$ for as-grown (standing) nanowires. Background corrected antibunching values are obtained following the analysis described in Ref. [26] using:

$$g^{(2)}_{cor}(0) = [g^{(2)}(0) - (1 - \rho^2)]/\rho^2, \quad (5)$$

where $\rho = S/(S + B)$ is the signal-to-total counts ratio, with $S$ denoting the signal and $B$ the background count rate. A value of ρ=0.92 for the lying nanowire and 0.72 for the standing one, respectively, was estimated from the emission spectrum recorded using the HBT detectors. The resulting $g_{cor}^{(2)}(0)$ of $0.06 \pm 0.04$ for lying and $0.32 \pm 0.15$ for standing nanowires suggests that the deviation from an ideal singe photon source can be attributed to the background contamination and finite temporal resolution of our photon counting setup. Indeed, for all studied single lying nanowires we obtained a $g_{cor}^{(2)}(0)$ value well below 0.5, which is the value expected for two independent single-photon emitters [25]. In the case of as-grown sample, the $g_{cor}^{(2)}(0)$ is currently limited by background emission from neighboring nanowire QDs. This background could be reduced by increasing the separation between nanowires.

In addition, continuous-wave photon correlation measurements allow determining the exciton emission rate $1/\tau_X$, which is a crucial parameter for high-speed operation of single-photon devices. This information is encoded in the temporal width of the antibunching dip $\tau_R$, which depends on $\tau_x$ and on the optical pump rate γ as [25]:

$$1/\tau_R = 1/\tau_X + \gamma, \qquad (6)$$

The coincidence histograms from a single lying nanowire were recorded for different pump powers and, as expected, we observed a linear increase of the antibunching rate $1/\tau_R$ with increasing the excitation power (cf. Fig. 5(c)). This linear trend extrapolates to $1/\tau_X = (0.84 \pm 0.10)\ ns^{-1}$ at vanishing pump power, giving an estimate of the exciton recombination time of $\tau_X = (1.19 \pm 0.14)\ ns$. This value is comparable with the exciton lifetime $\tau_X = (1.34 \pm 0.03)\ ns$ measured by time-resolved PL, as shown in Fig. 5(d). Such high emission rates suggest that these nanowire heterostructures are suitable for the realization of on-demand SPE operating at high-frequencies.

## 4. CONCLUSIONS

In conclusion, we have demonstrated PA-MBE-grown SPEs based on InGaN/GaN disk-in-a-wire nanostructures. The employed SAG technique allows controlling the location of these nanostructures, which form ordered hexagonal arrays. Such nanowire heterostructures can be easily transferred to foreign substrates, which facilitates their integration with other components for on-chip quantum information processing. In addition, epitaxial growth by MBE permits to tune the emission energy of the embedded InGaN nano-disks over a wide spectral range. The QD-like centers responsible for the emission of antibunched photons are located at the apexes of InGaN regions. Optical studies performed on these emission centers show narrow and highly linearly polarized PL lines. Their non-classical character is probed by HBT measurements. The values of the antibunching dip are well below the two-photon threshold of 0.5, therefore proving single photon emission. The short radiative life time of the emitted photons of ~1 ns, obtained independently from time-resolved PL and linear excitation power dependence of the antibunching rate, makes these SPEs suitable for high-frequency applications up to a GHz range.

**Acknowledgements**
This work was supported by the Spanish MINECO under contracts MAT2011-22997 and MAT2014-53119-C2-1-R, by CAM under contract S2009/ESP-1503 and by the FP7 ITN Spin-Optronics (237252). E.C. acknowledges the FPI grant (MINECO). The authors thank David Lopez Romero for the help in metal mask fabrication and Dr. Ana Bengoechea-Encabo and Steven Albert for the help in lithographic mask preparation and MBE growth. M.M. F.B. P.V. and J.C. thank the German Research Foundation DFG for financial support (research program INST 272/148-1) and the collaborative research center SFB 787 „Semiconductor Nanophotonics: Materials, Models, Devices".